  \documentclass{cambridge6A}
  \usepackage{natbib}
  \usepackage{aas_macros}
  \usepackage{amsmath}

\usepackage{graphicx}
\usepackage{amsmath}

\def\vv{{\boldsymbol v}}

\def\vxi{{\boldsymbol\xi}}
\newcommand\NoIndent[1]{%
  \begingroup
  \par
  \parshape0
  #1\par
  \endgroup
}

\begin{document}





\chapter{Sunquakes: helioseismic response to solar flares}
\vspace*{-3cm}
\begin{center}
{\normalsize Alexander G. Kosovichev\\ Big Bear Solar Observatory, NJIT\\ sasha@bbso.njit.edu}
\end{center}
\section*{Abstract}
Sunquakes observed in the form of expanding wave ripples on the
surface of the Sun during solar flares represent packets of acoustic waves
excited by flare impacts and traveling through the solar interior. The excitation impacts strongly correlate with the impulsive flare phase, and are caused by the energy and momentum transported from the energy release sites. The flare energy is released in the form of energetic particles, waves, mass motions, and radiation. However, the exact mechanism of the localized hydrodynamic impacts which generate sunquakes is unknown.  Solving the problem of the sunquake
mechanism will substantially improve our understanding of the flare physics. In addition, sunquakes offer a unique opportunity for studying the interaction of acoustic waves with magnetic fields and flows in flaring active regions, and for developing new approaches to helioseismic acoustic tomography.

\section{Introduction and overview}
\label{introduction}

\begin{figure}
\begin{center}
\includegraphics[width =\textwidth]{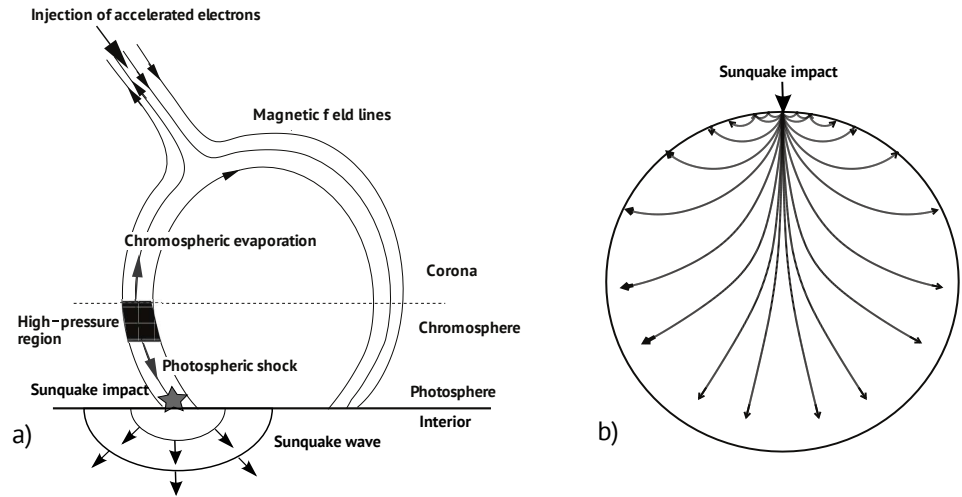}
\end{center}
\caption{a) Illustration of a sunquake mechanism in the `thick-target' model of solar flares: high-energy particles injected into a magnetic flux tube heat an upper chromospheric layer, creating a high-pressure region which produces `chromospheric evaporation' and a shock traveling to the photosphere and causing a sunquake impact. b) Illustration of ray paths of acoustic waves excited by an impulsive source on the solar surface. The waves travel into the deep interior and are reflected back to the surface, producing sunquake ripples.
 \label{fig1}}
\end{figure}

     Solar flares represent a process of rapid transformation of
the magnetic energy of active regions into the kinetic energy of charged
particles, plasma flows and heating of the solar atmosphere and corona.  The primary energy release during the flares is believed to occur in the corona as a result of magnetic reconnection.  It is generally believed that most of the energy
released by the reconnection goes directly and indirectly (via plasma waves) to
acceleration of electrons and protons which are injected into flaring loops (Fig.~\ref{fig1}$a$). Most of the observed radiation  is produced either directly by
these particles or indirectly through energization of the background plasma.

It was suggested long ago \citep{Wolff1972} that flares may cause acoustic waves traveling through the solar interior, similar to the seismic waves in the Earth. Because the sound speed increases with depth, the waves are reflected in the
deep layers of the Sun (Fig.~\ref{fig1}$b$), and then appear on the surface, forming expanding rings, similarly to ripples on the water surface. Theoretical modeling
\citep{Kosovichev1995} predicted that the speed of the expanding
seismic ripples increases with distance because the distant waves
propagate into the deeper interior where the sound speed is higher.

\begin{figure}
\begin{center}
\includegraphics[width =0.9\textwidth]{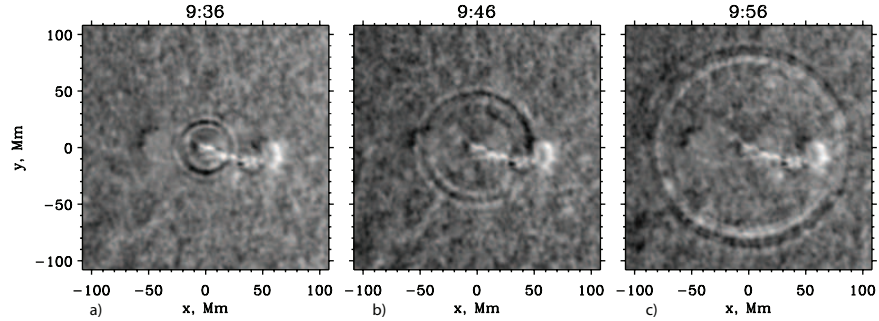}
\end{center}
\caption{\textit{A sequence of enhanced Doppler-velocity images of the first sunquake observed on July 9, 1996, with the Michelson Doppler Imager on SOHO spacecraft. The sunquake is observed on the solar surface as circular expanding wave originated from an impact of a X2.6 solar flare. Dark color shows upflows, and white color shows downflows. The noisy background is caused by solar granulation. Bright permanent features are flows associated with magnetic field of the active region. The sunquake wave signal in these images is enhanced by a factor of 4.}
 \label{fig2}}
\end{figure}

\begin{figure}
\begin{center}
\includegraphics[width =0.85\textwidth]{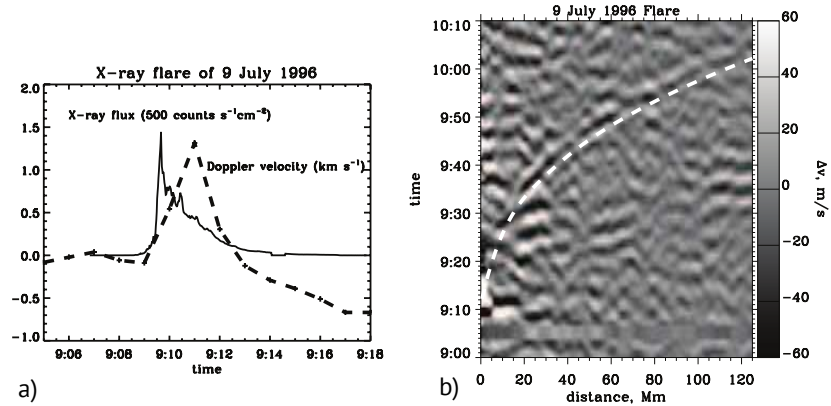}
\end{center}
\caption{a) The hard X-ray flux of the July 9, 1996, flare (solid) and the Doppler velocity signal of the localized flare impact which generated a sunquake; b) a time-distance diagram of the sunquake wave, obtained by azimuthally averaging and stacking the Doppler-velocity images obtained with 1-min cadence. The strong variations of velocity at about 9:10 UT represent the initial impact. The white dashed curve shows a theoretical time-distance relation of acoustic waves, calculated in a ray-path approximation.
 \label{fig3}}
\end{figure}

First observations of the seismic waves caused by the X2.6 flare of 9 July,
1996,  proved these predictions \citep[Fig.~\ref{fig2};][]{Kosovichev1998}.
These observations also showed that the source of the seismic response was a
strong shock-like compression wave propagating downwards in the photosphere.
This led to a suggestion that the seismic response can be explained in terms of so-called the hydrodynamic ``thick-target'' model (Fig.~\ref{fig1}$a$). In this model  \citep[e.g.][]{Kostiuk1975, Livshits1981, Fisher1985, Kosovichev1986}, a beam of high-energy particles, accelerated in the corona, heats the solar chromosphere, resulting in  a strong compression of the lower chromosphere. This  compression produces a chromospheric eruption (`evaporation') and a downward-propagating shock wave which hits the solar surface and causes a seismic response. This shock is observed in the SOHO/MDI Dopplergrams as a localized large-amplitude velocity impulse of about 1~km/s, immediately after the hard X-ray impulse produced by high-energy electrons (Fig.~\ref{fig3}$a$). This velocity impulse represents the initial hydrodynamic impact exciting the seismic waves. In addition,  it was found that the seismic waves had a significant anizotropic quadrupole component. However, the seismic wave front was almost circular (Fig.~\ref{fig2}).

Figure~\ref{fig3}$b$ shows a time-distance diagram for the spherically symmetrical component of the sunquake wave. The diagram is obtained by azimuthally averaging the Doppler velocity signal centered at the initial impact location. In this diagram, the initial impact started at about 9:10~UT, and the seismic wave is displayed as a narrow ridge with a characteristic slope, which follows the theoretical time-distance relations for acoustic waves traveling in the unperturbed conditions of the quit Sun. In this case, the theoretical relation is calculated for a standard solar model using the acoustic ray-path approximation \citep[e.g.][]{Kosovichev2011b}. While the appearance of sunquakes is similar to water ripples their physical properties are quite different. In particular, as follows from Fig.~\ref{fig3}$b$ the surface speed of sunquake waves rapidly increases with the travel distance. Immediately after the initiation the wave speed is about 10 km/s, which is close to the surface speed of sound, and then it rapidly increases to $\sim 100$~km/s. The apparent surface speed increases because the acoustic waves travel through the solar  interior where the sound speed is much higher than on the surface, and more distant waves travel through deeper interior (see Fig.~\ref{fig1}$b$).

 Subsequent observations of solar flares made
by the Michelson Doppler Imager (MDI) instrument on the NASA-ESA
mission SOHO \citep{Scherrer1995} did not show noticeable sunquake signals even for X-class flares until the big flares of October 2003. This search was carried out by calculating an``egression'' power for high-frequency acoustic waves during the
flares \citep{Donea1999}. Thus, it was suggested that sunquakes are a rather rare phenomenon on the Sun, which occur only under some special conditions. Surprisingly, seven years after the first event several flares did show strong ``egression'' signals indicating new potential sunquakes \citep{Donea2005}. Of course, detection of sunquakes depends on their amplitude relative to the background solar noise. Presumably, all flares generate some seismic response, but if the amplitude is not high enough the waves may be lost in the background noise.

The helioseismology observations with MDI discovered several powerful events in 2003-2005 during the declining phase of Solar Cycle 22  \citep{Kosovichev2006b,Donea2011}. Perhaps, the most powerful event of the cycle was on January 15, 2005 (Sec.~\ref{thick-target}). The observations revealed that in most events the helioseismic response is anisotropic, and that the wavefront shape may significantly deviate from circular. These effect are mostly associated with the sunquake source extension along flare ribbons and also with the source motion which can reach supersonic speeds \citep{Kosovichev2006a}. The source motion is often related to expanding flare ribbons, and probably reflects properties of magnetic reconnection and energy release in the solar atmosphere \citep{Sturrock1966a}. Recent observations with the Helioseismic and Magnetic Imager instrument on board Solar Dynamics Observatory \citep{Scherrer2012} show that sunquakes are much more common events than it was thought before, and that they can be frequently observed for moderate-class flares \citep{Kosovichev2012a}.

In this article, I present a brief overview of the basic theory and observational properties of sunquakes, discuss some recent results.

\section{Theory of the helioseismic response}
\label{theory}

Generally, the helioseismic response is caused by the energy and momentum transfer from a place of magnetic energy release in the solar atmosphere to the surface. Observations of Doppler velocities clearly reveal localized impacts in the low atmosphere and photosphere, which represent sources of sunquake events. Without such impacts sunquakes are not observed. Therefore, a theory of sunquakes must include two parts: 1) calculations of the hydrodynamic impact for a flare model assuming some mechanism of the energy release; 2) calculations of the helioseismic waves produced by this impact. So far, the theory is developed for a relatively simple hydrodynamic `thick-target' flare model, and the helioseismic response is calculated in the framework of a standard solar model and also for magnetostatic sunspot models.

\begin{figure}
\begin{center}
\includegraphics[width =0.9\textwidth]{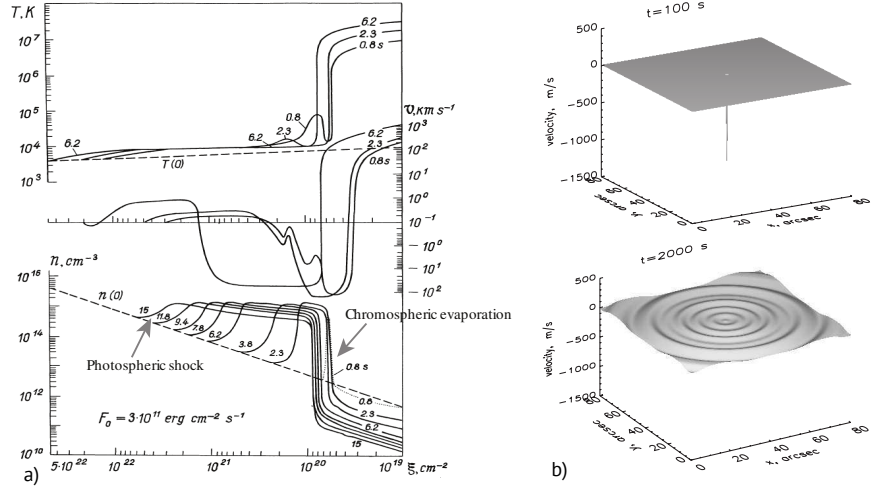}
\end{center}
\caption{a) Numerical simulations of the thick-target model of solar flares in a vertical magnetic flux tube with a constant cross-section: temperature, velocity and density profiles for different moments of time. The chromospheric heating and dynamics were initiated by a 10-sec injection of high-energy electrons at the top of the tube with the maximum energy flux of $3\times 10^{11}$ erg$\,$cm$^{-2}\,$sec$^{-1}$ with a power-law energy spectrum with the low-energy cut-off at 20~keV. The plasma compression behind the shock moving towards the photosphere can reach a factor of $\sim 100$ because of the radiative energy losses. This region can be a source of white-light emission. b) Theoretical model of sunquakes showing surface velocity perturbations at $t=200$ and 2000 s, produced by a point momentum source initiated at $t=0$.
 \label{fig4}}
\end{figure}

The hydrodynamic `thick-target' model was first developed by \citet{Kostiuk1975}, and calculated in various detail by many authors. In this model (schematically illustrated in Figure~\ref{fig1}$a$)  high-energy particles accelerated in the corona are injected into a magnetic flux tube.  They heat the upper chromospheric layers to high temperature, creating a high-pressure region which expands producing `chromospheric evaporation' and a shock traveling towards the photosphere. Numerical simulations of the flare dynamics in a one-dimensional magnetic flux tube model revealed that due to the intense radiative energy losses this shock can compress the plasma by a factor of $\sim 100$ to very high density \citep[Fig.~\ref{fig4}$a$;][]{Kosovichev1986}, and, thus, transfer substantial momentum into the lower atmosphere. The compressed relaxation zone behind the shock front can be also a source of the impulsive continuum emission observed in flares \citep{Livshits1981}. Quantitative details of the hydrodynamic `thick-target' model are still being developed \citep{Allred2005,RubiodaCosta2011}, and currently it is unclear if this shock can explain the impact sources observed in the Doppler shift of photospheric absorption lines.

The helioseismic response to a localized source can be calculated by using the standard stellar pulsation theory \citep[e.g.][]{Unno1989}.
A solution of the non-radial stellar pulsation equation written in symbolic form for a displacement vector $\vxi$:
   \begin{equation}
   \frac{\partial^2\vxi}{\partial  t^2}+{\cal L}\vxi=0
   \end{equation}
 for initial conditions: $\vv(r,\theta,\phi,0)=\vv_0(r,\theta,\phi) $
 can be obtained in terms of normal mode eigenfunctions, $\vxi_{nlm}$:
  \begin{equation}
  \vv(r,\theta,\phi,t)=\sum_{nlm}\frac{\langle\vxi_{nlm}^*\cdot\vv_0\rangle}
 {\langle\vxi_{nlm}^*\cdot\vxi_{nlm}\rangle}
 \vxi_{nlm}\cos(\omega_{nlm}t)e^{-\gamma_{nlm}t},
 \end{equation}
  where the angular
 brackets mean the integration over the solar mass. Assuming that the
 impact is localized in a very small volume at the surface and
 calculating the integrals one can obtain for the radial component of the
 oscillation velocity at the surface:
 \begin{equation}
  v_r(R,\theta,\phi,t)=\sum_{nlm}\frac{P_0}{M_\odot I_{nl}}
 (2\ell+1)P_\ell(\cos\theta)\cos(\omega_{nl}t)e^{-\gamma_{nl}t},
 \end{equation}
 where $P_0$ is the total momentum of the impact, $M_\odot$ is the
 solar mass, $I_{nl}$ is the mode inertia, $P_\ell$ is the Legendre
 polynomial, $\omega_{nl}$ are the mode eigenfrequencies, and
 $\gamma_{nl}$ are their damping times. The denominator, $M_\odot
 I_{nl}$, is often called `mode mass'. Thus, the mode amplitude
 excited by the impulsive point source is equal to the total momentum
 divided by the mode mass and multiplied by a geometrical factor
 $(2\ell+1)P_\ell$. The theoretical sunquake surface velocity signals are illustrated in Fig.~\ref{fig4}$b$.

\begin{figure}
\begin{center}
\includegraphics[width =0.9\textwidth]{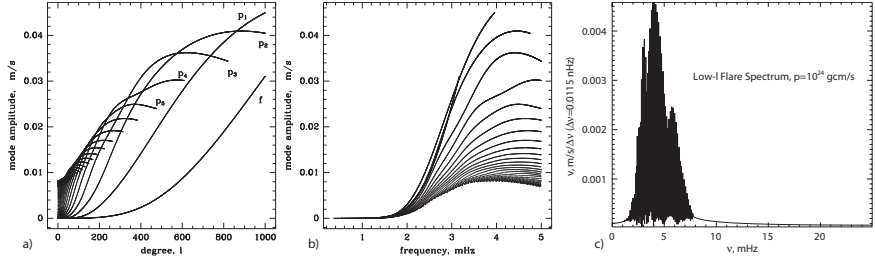}
\end{center}
\caption{a) Theoretical amplitudes of solar acoustic (p) and surface gravity (f) modes of various angular degree $\ell$, excited by the point source; b) the mode amplitudes as a function of oscillation frequency; c) the spectrum of low-degree modes ($\ell =0,1,2$) for observations of the Sun as a star. Most of the oscillation power of sunquakes is concentrated around the acoustic cut-off frequency at $\sim 5$ mHz.
 \label{fig5}}
\end{figure}

The spectrum of the excited oscillation modes is shown Fig.~\ref{fig5}$a,b$. For
the total momentum $10^{24}$ g\,cm/s, the calculated total amplitude corresponds to the maximum amplitude observed in the sunquake events. This means that the total momentum of the flare impact does not exceed $10^{24}$ g\,cm/s. Then using the same solution one can calculate the amplitudes of low-$\ell$ modes excited by this impact. The velocity spectrum of $\ell=0-2$ modes is shown in Fig.~\ref{fig5}$c$. The maximum amplitude of these global modes does not exceed 0.4 cm/s \citep{Kosovichev2009}. This is about 100 times smaller than
the amplitude of stochastically excited low-$\ell$ modes. The maximum amplitude of the flare-excited modes is in the frequency range of 4-5 mHz. These model results are inconsistent with the suggestion of \citet{Karoff2008} that they observed global sunquake oscillations. Perhaps, the high-frequency excess in intensity variations was caused not by oscillations but some other fluctuations associated with solar flares.

The helioseismic response in inhomogeneous sunspot regions can be studied only numerically. An example of such numerical simulations is presented in Sec.~\ref{magnetic}.

\section{Properties of sunquake sources}
\label{properties}

\subsection{Comparison observations with the thick-target flare model}
\label{thick-target}

\begin{figure}[t]
\begin{center}
\includegraphics[width =0.9\textwidth]{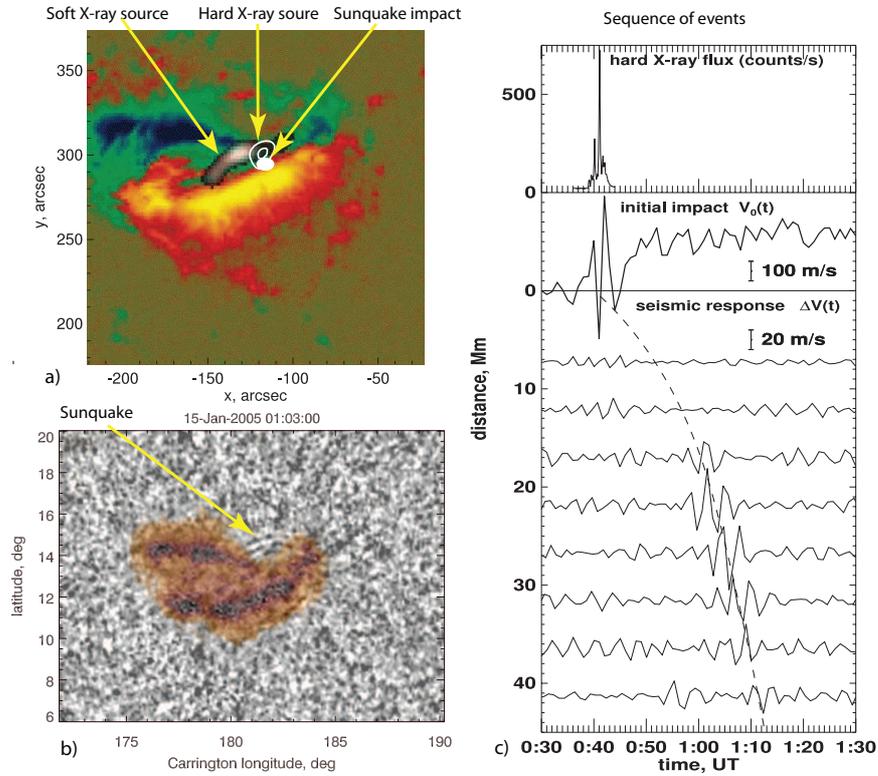}
\end{center}
\caption{Sunquake of X1.2 flare of January 15, 2005: a) location of the initial sunquake impact (white dot) on the SOHO/MDI magnetogram (color background), soft X-ray (gray-scale structure), and hard X-ray (contour lines) emissions observed by RHESSI; b) Doppler-velocity image showing the wave front (gray) overlaid with the corresponding image of sunspots (brown); c) the sequence of the sunquake events showing the hard X-ray impulse, and Doppler-velocity signals at the initial impact, and various distances from the impact point. The dashed curve shows the theoretical time-distance relation.
 \label{fig6}}
\end{figure}

Simultaneous observations of sunquake events using Doppler shift data and X-ray and gamma-ray flare emissions are used for investigation of the relationship between the flare-accelerated particles and the sunquake impacts in the photosphere. In particular, observations of X1.2 flare of January 15, 2015, showed  a good agreement with the `thick-target' flare model \citep{Kosovichev2006,Kosovichev2006b}. Figure~\ref{fig6}$a$ shows the soft (grey loop-like structure) and hard X-ray (white contours) emissions, as well as the location of the initial photospheric impact  (white spot) overlaid on the magnetogram. Figure~\ref{fig6}$b$ shows a superposition of the Doppler velocity and  continuum intensity images. Evidently, this event provides a nice example of the thick-target model configuration: the hard X-ray emission is observed at the footpoint of a magnetic loop filled with dense high-temperature plasma,  and the sunquake impact is observed just beneath the hard X-ray source. However, the sequence of events (Fig.~\ref{fig6}$c$) casts doubts in this model scenario. The photospheric impact started simultaneously or even earlier the hard X-ray peak (top two panel), whereas the thick-target model predicts a delay between the hard X-ray emission and the photospheric impact. In this model, the hard X-ray emission (bremsstrahlung) is produced by Coulomb scattering of high-electrons in the upper chromosphere. The upper chromosphere is heated during this process, and then expands causing the chromospheric evaporation and the photospheric shock. This means that there must be a significant delay between the hard X-ray impulse and the photospheric signal, roughly corresponding to a hydrodynamic response time of about 100-200 sec.  However, the observations do not show this delay.

\begin{figure}
\centering
\includegraphics[width=\linewidth]{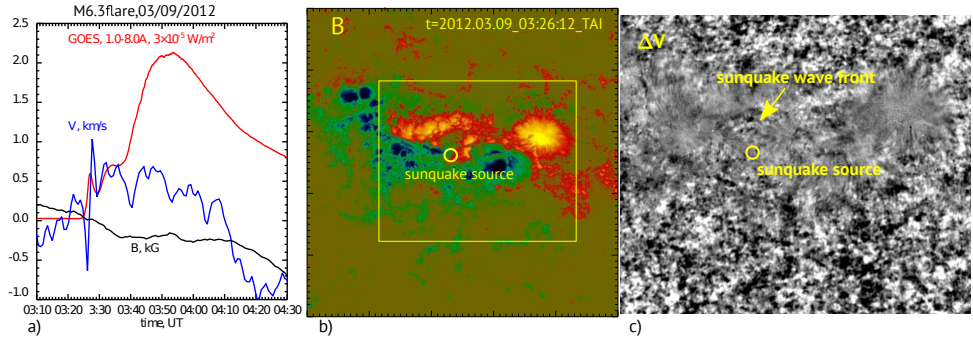}
\caption{Sunquake event produced in the pre-heating phase M6.3 flare of March 9, 2012: a) time profiles of the total soft X-ray flux (1.0-8.0 \AA) from GOES satellite (red), the photospheric velocity (blue) and magnetic field at the flare impact; b) the impact location  on the HMI magnetogram; c) the Doppler velocity variations showing the sunquake wave front in the area indicated on the magnetogram by a rectangle.}
\label{fig7}
\end{figure}

The discrepancy between the model predictions and observations became particularly apparent with the start of sunquake observations from Solar Dynamics Observatory. The first sunquake event detected by the HMI instrument on SDO showed that the hydrodynamic impact occurred at the very beginning of the flare impulsive phase and that the hard X-ray flux was rather low \citep{Kosovichev2011}. This discrepancy was particularly pronounced in  M6.3 flare of March 9, 2012, in which the photospheric impact at $\sim 3:27$ UT was more than 10 min prior the main impulsive phase, and was associated with a small peak of the soft X-ray emission \citep[Fig.~\ref{fig7}$a$][]{Kosovichev2012a}. Analysis of the HMI Doppler images and magnetograms showed that the impact was located in a region of relatively weak magnetic field, near a magnetic neutral line (Fig.~\ref{fig7}$b$). Variations of the magnetic field were observed after the velocity impact. This rules out the hypothesis of the magnetic origin of the impact due to an impulsive Lorentz force (so-called `McClymont jerk') \citep{Hudson2008}.  Observations of this flare with the Extreme-ultraviolet Imaging Spectrometer (EIS) on Hinode spacecraft revealed strong chromospheric evaporation characterized by 150-200 km/s upflows observed  at the same time as the photospheric impact in multiple locations in multi-million degree spectral lines of flare ions \citep{Doschek2013}. The detection of the chromospheric evaporation is consistent with the `thick-target' model scenario. However, the early timing of the impact indicates that the photospheric impact was probably caused by the energy release in the low atmosphere, and not by the energy transport from the high corona as suggested by the model. These observations may lead to the paradigm change in the theory of solar flares.

\subsection{Anisotropy of wavefront and source motion}
\label{anisotropy}

\begin{figure}
\begin{center}
\includegraphics[width =0.85\textwidth]{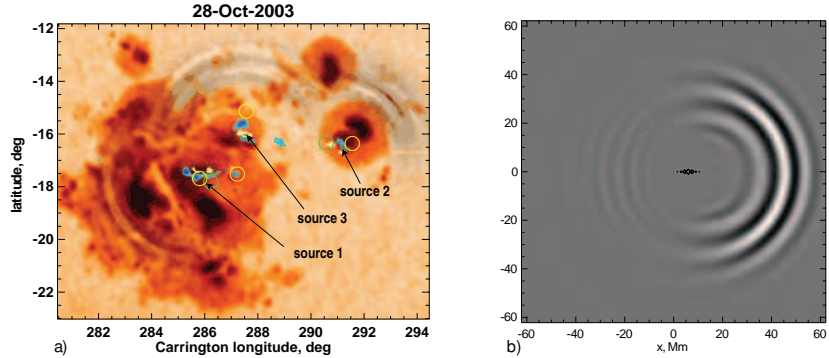}
\end{center}
\caption{a) Wavefronts of three sunquake events observed by the MDI instrument after the X17 flare of October 28, 2003, at 11:37 UT). The background is the corresponding MDI continuum intensity image; yellow and blue patches show down and up Doppler velocity signals with the amplitude stronger than 1 km/s, observed during the impulsive phase at 11:06 UT; yellow and green circles show locations of the hard X-ray (50-100 keV) and 2.2 MeV gamma-ray sources observed by RHESSI; b) Theoretical model of sunquakes with a moving impact source, which explains the observed anisotropy of sunquakes. The point impulsive source is moving in the $x$ direction with the constant speed of 25 km/s. Its strength as a function of time has a Gaussian shape with FWHM of 3 min. The locations of the source are shown by black diamonds at the center, the size of which is proportional to the source strength.
 \label{fig8}}
\end{figure}

A characteristic feature of the seismic response in this  and most  other flares is
anisotropy of the wave front: the observed wave amplitude is much
stronger in one direction than in the others. In particular, all three
seismic waves excited during the X17 flare of October 28, 2003, had the
greatest amplitude in the direction of the expanding flare ribbons (Fig.~\ref{fig8}$a$).
Thus, the wave anisotropy was attributed to the moving sources of the
hydrodynamic impact, which was located in the flare ribbons
\citep{Kosovichev2006b, Kosovichev2006a}. 
The motion of flare ribbons is often interpreted as a result of the magnetic  reconnection processes in the corona. When the reconnection region
moves up it involves higher magnetic loops, the footpoints of which
are located further apart. A sequence of energy injection events in the reconnecting loops create the effect of apparent source motion. Of course, there might be other reasons for the wave front anisotropy, such as inhomogeneities in temperature, magnetic field, and plasma flows. However, the source motion seems to be the primary factor.

In such cases, the seismic wave is generated not by a single impulse  but by a series of impulses moving on the solar surface. The seismic effect of the moving source can be easily calculated by convolving the wave Green's function with a time-dependent moving source function. The results of these calculations show a strong anisotropic wavefront (Fig.~\ref{fig8}$b$), qualitatively similar to the observations \cite{Kosovichev2007}. Curiously, this effect is quite similar to the anisotropy of seismic waves on Earth, when the earthquake rupture moves along the fault. Thus, taking into account the effects of multiple injections of accelerated particles in a realistic 3D magnetic geometry is very important for sunquake theories.

\subsection{Interaction of sunquake waves with magnetic field of sunspots}
\label{magnetic}

\begin{figure}
\centering
\includegraphics[width=0.95\linewidth]{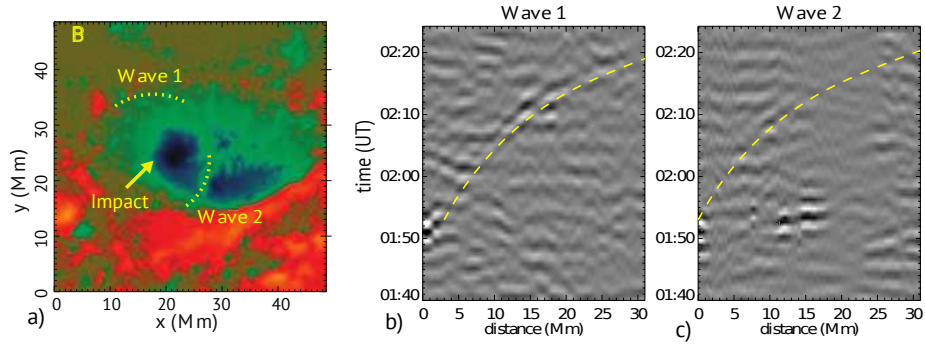}
\caption{Observations of sunquakes of X2.2 flare of February 15, 2011: a) SDO/HMI magnetogram of AR 11158 (red - positive polarity; blue - negative polarity) and locations of the initial photospheric impact (arrow), and two wave fronts: \textit{Wave 1} traveling through a weak-field region, and  \textit{Wave 2} traveling  through the sunspot umbra with strong magnetic field; panels b) and c) show time-distance diagrams of these wave fronts  illustrating that \textit{Wave 2} traveled slower and had a lower amplitude than \textit{Wave 1} (dashed curves show the theoretical time-distance relation).}
\label{fig9}
\end{figure}

Understanding of magnetic field effects on propagation of acoustic waves is of particular interest for helioseismology because this opens new diagnostics of subsurface structure and dynamics of sunspots and active regions \citep{Kosovichev2012}.  Local helioseismology techniques are based on measurements of perturbations of wave travel times and oscillation frequency shifts, which require long  time series (typically 8-24 hours) because of the stochastic nature of solar oscillations. Also, because the  distribution of wave excitation sources in active regions is not uniform the interpretation of cross-covariance functions and power spectra is not straightforward \citep[e.g.][]{Zhao2012a}. Sunquakes provide direct view of the wave fronts traveling through various areas of active regions with different magnetic properties and flow patterns. In combination with numerical modeling of MHD waves in sunspot models this provides valuable information about wave interaction with magnetic fields. These studies are in an initial stage, and here I present two observational and theoretical examples.

Figure~\ref{fig9}$a$ illustrates propagation of sunquake waves during X2.2 flare of February 15, 2011 \citep{Kosovichev2011}. The sunquake impact source was located in the sunspot penumbra close to the umbra-penumbra boundary. One part of the sunquake ($Wave 1$) traveled outside the spot through the penumbra whereas the other part ($Wave 2$) traveled through the sunspot umbra (dark blue region on the magnetogram). The time-distance diagrams of these waves (Fig.~\ref{fig9}$b,c$) show that $Wave 2$ traveled slower and had lower amplitude than $Wave 1$. This travel time delay for waves traveling through sunspots confirms the previous results obtained by time-distance helioseismology for relatively short travel distances \citep{Kosovichev2000}. However, it is important to note that the travel-time variations strongly depend on the frequency contents (bandwidth and mean frequency) of the acoustic wave packets.
\begin{figure}
\centering
\includegraphics[width=\linewidth]{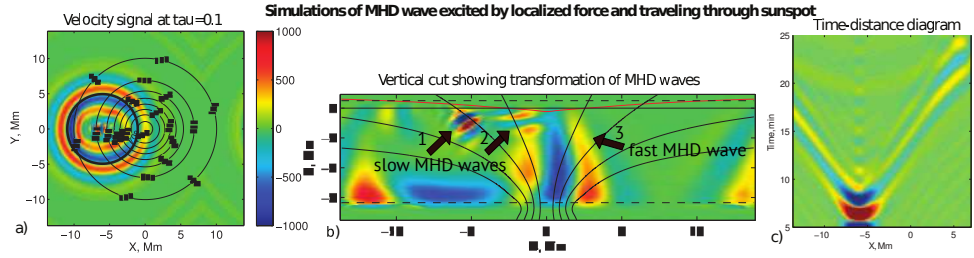}
\caption{Numerical 3D simulations of MHD waves excited by a localized impulsive force and traveling through a sunspot model: a) snapshot of the z-component of the velocity in the upper photosphere (optical depth $\tau=0.1$. The circular contours show magnetic field strength in Gauss; b) vertical cut of the simulated sunquake wave interacting with magnetic field. The helioseismic wave below the surface is a mixture of fast MHD and slow MHD waves; c) time-distance sunquake diagram of the photosphetic signal. }
\label{fig10}
\end{figure}

The origin of the amplitude decrease in sunspots is currently unknown. This can be due to several effects, including absorption of wave energy, wave scattering, transformation into different modes, changes in the acoustic cut-off frequency, etc.
Numerical simulations of sunquakes for sunspot models with different physical properties can help us to disentangle these effects and investigate their role. An example of such simulations  \citep{Parchevsky2009,Parchevsky2010} performed for a self-consistent magnetostatic sunspot model of \citet{Khomenko2008}  is shown in Fig.~\ref{fig10}. In these simulations the flare impact was modeled by a localized point-like impulsive vertical force at the outer edge of the sunspot model. The time profile of the impulse was chosen to model the frequency spectrum similar to the solar spectrum. One part of the excited wave traveled outside the magnetic structure, and the other part traveled through the structure. The amplitude reduction of the second wave front is apparent. The vertical cut in Fig.~\ref{fig10}$b$ shows that the impact force excited two types of MHD waves: a fast MHD wave (arrow 3) which formed circular sunquake-type ripples at the surface and a slow MHD wave (left arrow 1) which traveled along the magnetic field lines into the interior, and thus did not affect the surface wave pattern. Similarly, our simulations showed that when the slow MHD wave  (indicated by arrow 2) is excited because of the wave transformation at the level where plasma $\beta\equiv 8\pi P/B^2 =1$ it travels along the field lines into the interior. This allows us to conclude the observed sunquake phenomenon in magnetic field regions represents fast MHD waves, and that the wave transformation does not play significant role. The simulations reproduce the amplitude reduction when the waves travels through the strong field region. It seems that this effect is related to the wave reflection from the $\beta=1$ layer, which in sunspots is below the photosphere level. After the wave passes through the spot its amplitude restores to the quiet-Sun values, as illustrated by the time-distance diagram in Fig.~\ref{fig10}$c$. This means that the amplitude reduction in sunspots is probably not due to wave absorption, but rather due to changes in the surface reflection properties.

\subsection{Comparison of impact sources with holography reconstruction}
\label{holography}

\begin{figure}
\begin{center}
\includegraphics[width =\textwidth]{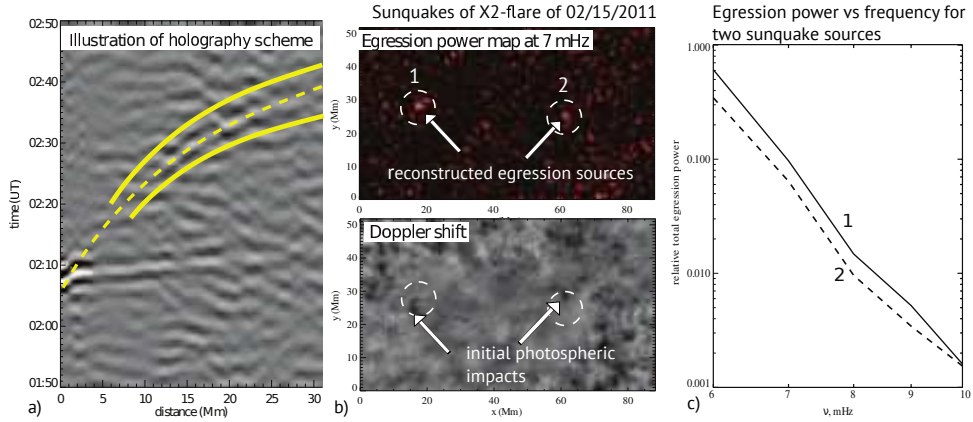}
\end{center}
\caption{Analysis of sunquakes using the holography approach: (a) illustration of the holography scheme, which integrates the wave-front signal and extrapolates it back in time to the origin assuming a standard time-distance relation (dashed curve) for solar acoustic waves; (b) top panel: egression power image of two sunquake events of X2-class flare of February 15, 2011 (dashed circles indicate the locations of the egression source), bottom panel: Doppler-velocity image showing the initial sunquake impacts as black point sources near the edges of the dashed circles; (c) the egression power as a function of frequency for the two events of the X2 flare (marked 1 and 2).
 \label{fig11}}
\end{figure}

There are two approaches for detection and investigation of sunquakes. The first approach is based on visual detection of the initial impacts and traveling sunquake waves (which are usually best visible $\sim 20$ min after the impact) in a series of images of Doppler-shift running differences processed through a high-pass filter in order to partially separate sunquake signals from convective noise, and also on the time-distance diagrams constructed for the impact locations. Generally, strong initial impacts in the photosphere with a speed of 1 km/s or greater are good indications of sunquakes. Such impacts are easily observed in Dopplergrams.The second approach, called `acoustic holography' \citep{Donea1999} attempts to reconstruct the initial sources by tracing the sunquake wave signals back in time to their origins, assuming that the wave propagation follows theoretical Green's function relations for point acoustic sources in the horizontally  uniform quiet-Sun models (see Fig.~\ref{fig11}$a$). The reconstruction is performed for different frequency filters for every point in the flaring region, and the results are plotted as `egression power maps'. Then these maps are visually inspected for spikes which are considered as candidates of sunquake sources. When these spikes coincide with localized flare indicators as sources of continuum emission they are identified as sunquakes. For strong events both approaches are equally reliable. However, when the flare perturbations are weak the time-distance approach may miss weak sunquake events while the holography reconstructions may lead to false positive identifications. In addition, because in the real Sun conditions the impact sources and the wave properties may be quite different from the theoretical Green's function. Therefore, the location and times of the reconstructed sources may differ from the initial flare impacts (Kosovichev \& Zhao, in preparation). This discrepancy is illustrated in Fig.~\ref{fig11}$b$, in which the top panel shows locations of two sunquakes source identified in the egression power map at 7 mHz, and the bottom panel presents a Doppler shift map, processed for the time-distance analysis, showing locations of the actual flare impacts. Evidently, there is a shift of about 5~Mm in these locations. This discrepancy has to be taken into account in sunquake studies.

\section{Challenges in understanding sunquakes}
\label{conclusion}

Since the first detection in 1996, sunquakes have been extensively studied using data from the SOHO and SDO space missions, and also from the ground-based network GONG. However, the mechanism of these events are their physical properties are still mysterious. Here, I summarize some of the primary questions, potential answers, problems and challenges.
\begin{itemize}
\item \textit{Why are sunquake events observed only in some flares?}
\NoIndent{\noindent New observations with the HMI instrument on SDO reveal that sunquakes are more common phenomenon that it was thought before. These observations showed that sunquake events may occur even for relatively weak flares of the X-ray class as low as M1, whereas in some strong X-class flares  the photospheric impacts are almost absent, and helioseismic response is not detectable.  A full statistical analysis of the HMI observations is not completed yet, and I can only make some preliminary conclusions. It appears that the helioseismic response is a characteristic of compact energy release events in the low atmosphere. Such energy release events usually are not accompanied by coronal mass ejections (CMEs), but generate coronal shocks (often observed by the SDO/AIA instrument). In flares, accompanied (and perhaps triggered) by CMEs the energy is often released in the high corona. Such flares may have a high X-ray class, but produce virtually no photospheric impact. My current conjecture is that sunquakes are more common for `compact' (`confined') flares than for `erruptive' (`dynamic') flares. This flare classification introduced by \citet{Pallavicini1977} and \citet{Svestka1986} is based on characteristics of X-ray and optical emissions and dynamics.  Complex flares (such as the X2.2 flare of February 15, 2011) may have both `compact' and `erruptive' components. The relationship between sunquakes and the two types of flares requires detailed statistical studies, and is certainly  very important for our understanding of the basic mechanisms of energy release in solar flares.}

\item \textit{Why are the initial photospheric impacts observed in the early impulsive phase?}
\NoIndent{\noindent The photospheric impacts observed at the very beginning of the flare impulsive phase or even during the pre-heating phase represent one of the greatest puzzles of sunquakes and challenge the standard flare model. The standard thick-target model (Sec.~\ref{thick-target}) predicts that the photospheric impact is a result of an expansion (accompanied by shocks) of a upper chromosphere region heated by high-energy electrons accelerated in the corona. In this scenario the photospheric impact should follow the hard X-ray impulse with a time delay of  $\sim 10^2$ sec, which is needed for the perturbations and shocks to propagate from the upper chromosphere to the photosphere. However, in many events the photospheric impact is observed at the beginning of the impulsive phase before the hard X-ray impulse reaches the maximum. This can be explained by two reasons: 1) the low-chromosphere plasma is directly heated by deeply penetrating particles, electrons or protons, which do not produce significant X-ray or gamma-ray emission; 2) the initial sunquake impacts are due to the flare-related dynamics, e.g. caused by the Lorentz force or by a flux-rope eruption, rather than high-energy particles. The first scenario seems to be more likely because of the initial sunquake impacts are often synchronised over very large distances, which can be achieved only by relativistic particles, and also because the speed of apparent motion of sunquake sources on the surface is often significantly higher than the local Alfven speed or sound speed.}

\begin{figure}
\centering
\includegraphics[width=0.9\linewidth]{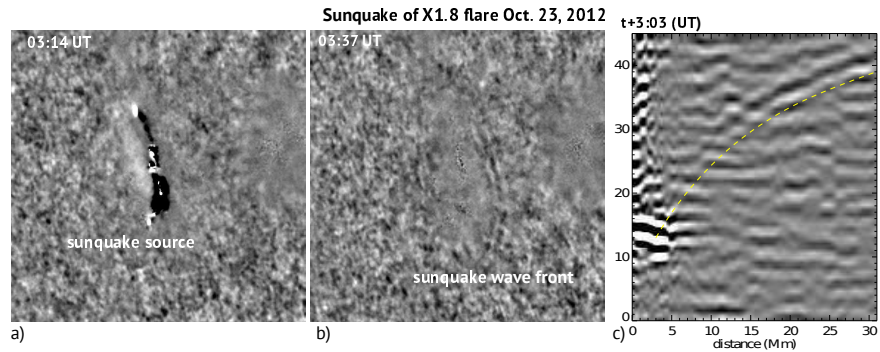}
\caption{One of the strongest events of Solar Cycle 24 observed during X1.8 flare of October 23, 2012: a) Doppler velocity variations at 03:14 UT, showing an unusually long impact source extended along the magnetic neutral line; b) the sunquake wave front at 03:37 UT; c) the sunquake time-distance diagram, showing the source motion at the beginning of the flare and sunquake ridge (yellow dashed curve is the theoretical relation).}
\label{fig12}
\end{figure}

\item \textit{What is the mechanism of the sunquake anisotropy and variations of wave amplitude?}
\NoIndent{\noindent Observations show that most sunquakes are highly anisotropic. The anisotropy is usually observed as a significant enhancement of the wave amplitude in one direction compared the amplitude in other directions. The extent of the amplitude enhancement can be very narrow or can extend over almost a half of the circle. Also, the wave front can be far from circular. For instance, in the greatest sunquake event observed so far in Solar Cycle 24 on October 23, 2012, during X1.8 flare, the sunquake wavefront had a very elongated elliptical shape with the strongest amplitude on the one side of the ellipse (Fig.~\ref{fig12}$b$). Such shape is due to a almost linear source (Fig~\ref{fig12}$a$) caused by a series of almost simultaneous impacts located in a flare ribbon parallel to the magnetic neutral line. Such extended sources are not uncommon, but, perhaps, this was the longest source observed with the space MDI and HMI instruments. Generally, there are two main sources of the wavefront anisotropy: 1) wave propagation through the inhomogeneous structure and flows of active regions; 2) rapid motion of the excitation source. Typically, when the waves travel through regions of strong magnetic field of sunspots their amplitude is reduced, but when the wave reemerges from sunspots the amplitude is restored to the quiet-Sun values. The numerical simulations presented in Sec.~\ref{magnetic} show that impulsive force or momentum impacts in magnetic regions excite fast and slow MHD waves. However, the slow wave travels along the magnetic field lines into the deep interior, and does not appear in the surface signal. Also, when the fast MHD wave reaches the region of plasma $\beta=1$ it is partially transformed in slow MHD wave, which also travels into the interior and does not affect the observed wave fronts. When the wave traveling through a sunspot emerges on the other side its amplitude is restored to the quiet-Sun values. The rapid source motion is a typical feature of sunquakes, and, likely, the primary cause of the wave amplification and anisotropy due to the interference effect (Sec.~\ref{anisotropy}). 
}

\item \textit{What causes the rapid motion of sunquake sources?}
\NoIndent{\noindent The sunquake source motion is observed as a sequence of individual point-like impacts \citep{Kosovichev2006a} or as a wave-like motion \citep{Kosovichev2011a}. In the first case, the apparent speed can reach 50-60 km/s, substantially exceeding the speed of sound or Alfven waves. The impact sources are usually located in the flare ribbons, and this motion corresponds to the ribbon expansion, but sometimes the impact sources can move along the ribbons \citep{Kosovichev2007}. The cause of the ribbon expansion is not fully understood. The current paradigm, based on the thick-target model, is that the flare ribbons represent a response of the low atmosphere to the precipitation of energetic particles accelerated in a magnetic reconnection region, and that the ribbons expands as the reconnection source moves higher up in the corona. The rapid source motion, particularly, in the form of a sequence of discrete impacts, clearly indicates that the sunquakes most likely represent a helioseismic response to a series of impacts of flare-accelerated  particles.}

\item \textit{What is the relationship between the flare accelerated particles and sunquake initiation?}
\NoIndent{\noindent Perhaps, this is the central question, which represents the importance of sunquakes for our understanding of solar flares. The spatial and temporal properties of the sunquake sources clearly indicate a close association with flare-accelerated particles, however, this relationship is not as straightforward as it follows from the flare thick-target model (Fig.~\ref{fig1}$a$). It is particularly puzzling that the sunquake impacts are observed at the very beginning of the impulsive phase, and that it seems that there is no clear correlation between the hard X-ray and gamma-ray emission of solar flares. One idea to solve this puzzle is that the particles that transport the flare energy into the low chromosphere are of relatively low energy and do not produce significant X-ray and gamma-ray emissions. Perhaps, these particles are accelerated in the low chromosphere, or penetrate through low-density magnetic flux tubes. These and other possibilities require detailed theoretical modeling and comparison with other observational effects, such as the transient variations of the photospheric line profiles, including rapid variations of magnetogram signals, and continuum emission \citep[e.g.][]{Kosovichev2001}.}

\item \textit{Can sunquake be cause by variations of magnetic field (Lorentz force) or by filament eruptions?}
\NoIndent{\noindent Alternative ideas to explain the localized impacts in the low atmosphere include effects of the Lorentz force due to the rapid restructuring of
magnetic field configuration, resulting from magnetic reconnection and 
dissipation \citep[e.g.][]{Hudson2008}. However, the observations do not show a correlation between the magnetic field variations and the sunquake impact sources \citep[e.g.][]{Kosovichev2011}. Also, quite often the impacts are observed in weak-field regions (e.g. see Fig.~\ref{fig7}). It was also suggested that the back reaction of erupting filaments (flux ropes) or CMEs might generate sunquake events \citep[e.g.][]{Zharkov2011}. In this case the synchronization of distant impact sources has to be with the Alfven speed, which is quite slow in the low atmosphere. This contradicts to the observations of almost simultaneous impacts over long distances (e.g. see Fig.~\ref{fig12}$a$). Also, in many sunquake cases flux rope eruptions are not observed. Therefore, it seems unlikely that the sunquake photospheric impacts and helioseismic waves are caused by these mechanisms.}


\item \textit{Can the seismic response flares be observed in global oscillation modes and on other stars?}
\NoIndent{\noindent The expanding sunquake waves represent high-degree p-modes, however, the theory (Sec.~\ref{theory}) predicts that the global low-degree modes are also excited. Their detection has
been reported from a statistical analysis of the correlation between
the flare soft X-ray signals (observed by the GOES satellites) and
the total solar irradiance measurements from the space observatory
SOHO \citep{Karoff2008}. However, there was no unambiguous detection of the whole-Sun oscillations caused by individual flare events. Indeed, the theory predicts that the amplitude of the low-degree modes of sunquakes is significantly lower than the amplitude of stochastically excited oscillations (Fig.~\ref{fig5}$c$) \citep{Kosovichev2009}. However, in the case of
significantly more powerful stellar flares the impact on the star's
surface can be much greater. This may lead to excitation of the global
low-degree oscillations to significantly higher amplitudes than on
the Sun. A preliminary study of the Kepler short-cadence
data provides indications of such `starquakes' \citep{Kosovichev2014}. However, a
statistical study for a large sample of stars and longer observing
intervals are needed for investigation of the `starquakes'.
}
\end{itemize}



\renewcommand{\refname}{Bibliography}
\bibliographystyle{cambridgeauthordate}

\begin{thebibliography}{39}
\expandafter\ifx\csname natexlab\endcsname\relax\def\natexlab#1{#1}\fi
\expandafter\ifx\csname selectlanguage\endcsname\relax
  \def\selectlanguage#1{\relax}\fi

\bibitem[\protect\citename{{Allred} {et~al.}, }2005]{Allred2005}
{Allred}, J.~C., {Hawley}, S.~L., {Abbett}, W.~P., and {Carlsson}, M. 2005.
\newblock {Radiative Hydrodynamic Models of the Optical and Ultraviolet
  Emission from Solar Flares}.
\newblock {\em \apj}, {\bf 630}, 573--586.

\bibitem[\protect\citename{{Donea}, }2011]{Donea2011}
{Donea}, A. 2011.
\newblock {Seismic Transients from Flares in Solar Cycle 23}.
\newblock {\em \ssr}, {\bf 158}, 451--469.

\bibitem[\protect\citename{{Donea} and {Lindsey}, }2005]{Donea2005}
{Donea}, A.-C., and {Lindsey}, C. 2005.
\newblock {Seismic Emission from the Solar Flares of 2003 October 28 and 29}.
\newblock {\em \apj}, {\bf 630}, 1168--1183.

\bibitem[\protect\citename{{Donea} {et~al.}, }1999]{Donea1999}
{Donea}, A.~C., {Lindsey}, C., and {Braun}, D. 1999.
\newblock {Helioseismic Holography - a Technique for Understanding Solar
  Flares}.
\newblock {\em Romanian Astronomical Journal}, {\bf 9}, 71.

\bibitem[\protect\citename{{Doschek} {et~al.}, }2013]{Doschek2013}
{Doschek}, G.~A., {Warren}, H.~P., and {Young}, P.~R. 2013.
\newblock {Chromospheric Evaporation in an M1.8 Flare Observed by the
  Extreme-ultraviolet Imaging Spectrometer on Hinode}.
\newblock {\em \apj}, {\bf 767}, 55.


\bibitem[\protect\citename{{Fisher} {et~al.}, }1985]{Fisher1985}
{Fisher}, G.~H., {Canfield}, R.~C., and {McClymont}, A.~N. 1985.
\newblock {Flare Loop Radiative Hydrodynamics - Part Six - Chromospheric
  Evaporation due to Heating by Nonthermal Electrons}.
\newblock {\em \apj}, {\bf 289}, 425.

\bibitem[\protect\citename{{Hudson} {et~al.}, }2008]{Hudson2008}
{Hudson}, H.~S., {Fisher}, G.~H., and {Welsch}, B.~T. 2008.
\newblock {Flare Energy and Magnetic Field Variations}.
\newblock {In:} {Howe}, R., {Komm}, R.~W., {Balasubramaniam}, K.~S.,
  and {Petrie}, G.~J.~D. (eds), {\em Subsurface and Atmospheric Influences on
  Solar Activity}.
\newblock Astronomical Society of the Pacific Conference Series, vol. 383, 221.

\bibitem[\protect\citename{{Karoff} and {Kjeldsen}, }2008]{Karoff2008}
{Karoff}, C., and {Kjeldsen}, H. 2008.
\newblock {Evidence That Solar Flares Drive Global Oscillations in the Sun}.
\newblock {\em \apjl}, {\bf 678}, L73--L76.

\bibitem[\protect\citename{{Khomenko} and {Collados}, }2008]{Khomenko2008}
{Khomenko}, E., and {Collados}, M. 2008.
\newblock {Magnetohydrostatic Sunspot Models from Deep Subphotospheric to
  Chromospheric Layers}.
\newblock {\em \apj}, {\bf 689}, 1379--1387.



\bibitem[\protect\citename{{Kosovichev}, }1986]{Kosovichev1986}
{Kosovichev}, A.~G. 1986.
\newblock {Simulating thermal and gasdynamic processes in solar-flare impulse
  phases}.
\newblock {\em Bulletin Crimean Astrophysical Observatory}, {\bf 75}, 6.

\bibitem[\protect\citename{{Kosovichev}, }2006a]{Kosovichev2006}
{Kosovichev}, A.~G. 2006a.
\newblock {Direct Observations of Acoustic Waves Excited by Solar Flares and
  their Propagation in Sunspot Regions}.
\newblock {In:} {J.~Leibacher, R.~F.~Stein, \& H.~Uitenbroek} (ed),
  {\em Solar MHD Theory and Observations: A High Spatial Resolution
  Perspective}.
\newblock Astronomical Society of the Pacific Conference Series, vol. 354, 154.

\bibitem[\protect\citename{{Kosovichev}, }2006b]{Kosovichev2006b}
{Kosovichev}, A.~G. 2006b.
\newblock {Properties of Flares-Generated Seismic Waves on the Sun}.
\newblock {\em \solphys}, {\bf 238}, 1--11.

\bibitem[\protect\citename{{Kosovichev}, }2006c]{Kosovichev2006a}
{Kosovichev}, A.~G. 2006c.
\newblock {Sunquake sources and wave propagation}.
\newblock {In:} {\em Proceedings of SOHO 18/GONG 2006/HELAS I, Beyond the
  spherical Sun}. \newblock ESA Special Publication, vol. 624, p.134.1.

\bibitem[\protect\citename{{Kosovichev}, }2007]{Kosovichev2007}
{Kosovichev}, A.~G. 2007.
\newblock {The Cause of Photospheric and Helioseismic Responses to Solar
  Flares: High-Energy Electrons or Protons?}
\newblock {\em \apjl}, {\bf 670}, L65--L68.

\bibitem[\protect\citename{{Kosovichev}, }2009]{Kosovichev2009}
{Kosovichev}, A.~G. 2009.
\newblock {Solar Oscillations}.
\newblock {In:} {Guzik}, J.~A., and {Bradley}, P.~A. (eds),
  {\em American Institute of Physics Conference Series}.
\newblock American Institute of Physics Conference Series, vol. 1170, 547--559.

\bibitem[\protect\citename{{Kosovichev}, }2011a]{Kosovichev2011b}
{Kosovichev}, A.~G. 2011a.
\newblock {Advances in Global and Local Helioseismology: An Introductory
  Review}.
\newblock {In:} {Rozelot}, J.-P., and {Neiner}, C. (eds), {\em
  Lecture Notes in Physics, Berlin Springer Verlag}.
\newblock Lecture Notes in Physics, Berlin Springer Verlag, vol. 832, 3-84.


\bibitem[\protect\citename{{Kosovichev}, }2011b]{Kosovichev2011a}
{Kosovichev}, A.~G. 2011b.
\newblock {First Sunquake of Solar Cycle 24 Observed by Solar Dynamics
  Observatory}.
\newblock {\em ArXiv e-prints}, 1102.3954.

\bibitem[\protect\citename{{Kosovichev}, }2011c]{Kosovichev2011}
{Kosovichev}, A.~G. 2011c.
\newblock {Helioseismic Response to the X2.2 Solar Flare of 2011 February 15}.
\newblock {\em \apjl}, {\bf 734}, L15.

\bibitem[\protect\citename{{Kosovichev}, }2012a]{Kosovichev2012}
{Kosovichev}, A.~G. 2012a.
\newblock {Local Helioseismology of Sunspots: Current Status and Perspectives}.
\newblock {\em \solphys}, {\bf 279}, 323--348.

\bibitem[\protect\citename{{Kosovichev}, }2012b]{Kosovichev2012a}
{Kosovichev}, A.~G. 2012b.
\newblock {Physics of Sunquakes Events Observed with SDO}.
\newblock {\em American Astronomical Society Meeting
  Abstracts}, vol. 220, 109.03.

\bibitem[\protect\citename{{Kosovichev}, }2014]{Kosovichev2014}
{Kosovichev}, A.~G. 2014.
\newblock Sunquakes and starquakes.
\newblock {In:} {Chaplin}~W.J., {Guzik}~J.A., {Handler}~G., and
  A., {Pigulski} (eds), {\em IAU Symposium 301 – Precision Asteroseismology:
  A Celebration of Scientific Opus of Wojciech Dziembowski},  vol. 301.
\newblock Cambridge Univ. Press, in press, eprint arXiv:1401.8036.

\bibitem[\protect\citename{{Kosovichev} and {Zharkova}, }1995]{Kosovichev1995}
{Kosovichev}, A.~G., and {Zharkova}, V.~V. 1995.
\newblock {Seismic Response to Solar Flares: Theoretical Predictions}.
\newblock {In:} {\em Helioseismology}.
\newblock ESA Special Publication, vol. 376, 341.

\bibitem[\protect\citename{{Kosovichev} and {Zharkova}, }1998]{Kosovichev1998}
{Kosovichev}, A.~G., and {Zharkova}, V.~V. 1998.
\newblock {X-ray flare sparks quake inside Sun}.
\newblock {\em \nat}, {\bf 393}, 317--318.

\bibitem[\protect\citename{{Kosovichev} and {Zharkova}, }2001]{Kosovichev2001}
{Kosovichev}, A.~G., and {Zharkova}, V.~V. 2001.
\newblock {Magnetic Energy Release and Transients in the Solar Flare of 2000
  July 14}.
\newblock {\em \apjl}, {\bf 550}, L105--L108.

\bibitem[\protect\citename{{Kosovichev} {et~al.}, }2000]{Kosovichev2000}
{Kosovichev}, A.~G., {Duvall}, Jr., T.~L., and {Scherrer}, P.~H. 2000.
\newblock {Time-Distance Inversion Methods and Results - (Invited Review)}.
\newblock {\em \solphys}, {\bf 192}, 159--176.

\bibitem[\protect\citename{{Kostiuk} and {Pikelner}, }1975]{Kostiuk1975}
{Kostiuk}, N.~D., and {Pikelner}, S.~B. 1975.
\newblock {Gasdynamics of a flare region heated by a stream of high-velocity
  electrons}.
\newblock {\em \sovast}, {\bf 18}, 590--599.

\bibitem[\protect\citename{{Livshits} {et~al.}, }1981]{Livshits1981}
{Livshits}, M.~A., {Badalian}, O.~G., {Kosovichev}, A.~G., and {Katsova}, M.~M.
  1981.
\newblock {The optical continuum of solar and stellar flares}.
\newblock {\em \solphys}, {\bf 73}, 269--288.

\bibitem[\protect\citename{{Pallavicini} {et~al.}, }1977]{Pallavicini1977}
{Pallavicini}, R., {Serio}, S., and {Vaiana}, G.~S. 1977.
\newblock {A survey of soft X-ray limb flare images - The relation between
  their structure in the corona and other physical parameters}.
\newblock {\em \apj}, {\bf 216}, 108--122.



\bibitem[\protect\citename{{Parchevsky} {et~al.}, }2010]{Parchevsky2010}
{Parchevsky}, K., {Kosovichev}, A., {Khomenko}, E., {Olshevsky}, V., and
  {Collados}, M. 2010.
\newblock {Numerical Simulation of Excitation and Propagation of Helioseismic
  MHD Waves in Magnetostatic Models of Sunspots}.
\newblock {\em ArXiv e-prints}, 1002.1117.

\bibitem[\protect\citename{{Parchevsky} and {Kosovichev},
  }2009]{Parchevsky2009}
{Parchevsky}, K.~V., and {Kosovichev}, A.~G. 2009.
\newblock {Excitation, Propagation and Conversion of Helioseismic MHD Waves in
  Strong Field Regions}.
\newblock {In:} {Dikpati}, M., {Arentoft}, T., {Gonz{\'a}lez
  Hern{\'a}ndez}, I., {Lindsey}, C., and {Hill}, F. (eds), {\em Solar-Stellar
  Dynamos as Revealed by Helio- and Asteroseismology: GONG 2008/SOHO 21}.
\newblock Astronomical Society of the Pacific Conference Series, vol. 416, 61.


\bibitem[\protect\citename{{Rubio da Costa} {et~al.}, }2011]{RubiodaCosta2011}
{Rubio da Costa}, F., {Zuccarello}, F., {Labrosse}, N., {Fletcher}, L.,
  {Proseck{\'y}}, T., and {Ka{\v s}parov{\'a}}, J. 2011.
\newblock {Solar flares: observations vs simulations}.
\newblock {Pages  182--184 of:} {Bonanno}, A., {de Gouveia Dal Pino}, E., and
  {Kosovichev}, A.~G. (eds), {\em IAU Symposium}.
\newblock IAU Symposium, vol. 274.

\bibitem[\protect\citename{{Scherrer} {et~al.}, }1995]{Scherrer1995}
{Scherrer}, P.~H., {Bogart}, R.~S., {Bush}, R.~I., {Hoeksema}, J.~T.,
  {Kosovichev}, A.~G., {Schou}, J., {Rosenberg}, W., {Springer}, L., {Tarbell},
  T.~D., {Title}, A., {Wolfson}, C.~J., {Zayer}, I., and {MDI Engineering
  Team}. 1995.
\newblock {The Solar Oscillations Investigation - Michelson Doppler Imager}.
\newblock {\em \solphys}, {\bf 162}, 129--188.

\bibitem[\protect\citename{{Scherrer} {et~al.}, }2012]{Scherrer2012}
{Scherrer}, P.~H., {Schou}, J., {Bush}, R.~I., {Kosovichev}, A.~G., {Bogart},
  R.~S., {Hoeksema}, J.~T., {Liu}, Y., {Duvall}, T.~L., {Zhao}, J., {Title},
  A.~M., {Schrijver}, C.~J., {Tarbell}, T.~D., and {Tomczyk}, S. 2012.
\newblock {The Helioseismic and Magnetic Imager (HMI) Investigation for the
  Solar Dynamics Observatory (SDO)}.
\newblock {\em \solphys}, {\bf 275}, 207--227.

\bibitem[\protect\citename{{Sturrock}, }1966]{Sturrock1966a}
{Sturrock}, P.~A. 1966.
\newblock {Model of the High-Energy Phase of Solar Flares}.
\newblock {\em \nat}, {\bf 211}, 695--697.

\bibitem[\protect\citename{{Svestka}, }1986]{Svestka1986}
{Svestka}, Z. 1986.
\newblock {On the varieties of solar flares}.
\newblock {In:} {Neidig}, D.~F. (ed), {\em The lower atmosphere
  of solar flares}; Proceedings of the Solar Maximum Mission Symposium, Sunspot,
  NM, Aug. 20-24, 1985 (A87-26201 10-92). Sunspot, NM, National Solar
  Observatory, 1986, p. 332-355.

\bibitem[\protect\citename{{Unno} {et~al.}, }1989]{Unno1989}
{Unno}, W., {Osaki}, Y., {Ando}, H., {Saio}, H., and {Shibahashi}, H. 1989.
\newblock {\em {Nonradial oscillations of stars}}, Tokyo Univ. Press.

\bibitem[\protect\citename{{Wolff}, }1972]{Wolff1972}
{Wolff}, C.~L. 1972.
\newblock {Free Oscillations of the Sun and Their Possible Stimulation by Solar
  Flares}.
\newblock {\em \apj}, {\bf 176}, 833.

\bibitem[\protect\citename{{Zhao} {et~al.}, }2012]{Zhao2012a}
{Zhao}, J., {Parchevsky}, K.~V., and {Kosovichev}, A.~G. 2012.
\newblock {Interaction of Helioseismic Waves with Sunspots: Observations and
  Numerical MHD Simulations}.
\newblock {In:} {Shibahashi}, H., {Takata}, M., and {Lynas-Gray},
  A.~E. (eds), {\em Progress in Solar/Stellar Physics with Helio- and
  Asteroseismology}.
\newblock Astronomical Society of the Pacific Conference Series, vol. 462, 277.

\bibitem[\protect\citename{{Zharkov} {et~al.}, }2011]{Zharkov2011}
{Zharkov}, S., {Green}, L.~M., {Matthews}, S.~A., and {Zharkova}, V.~V. 2011.
\newblock {2011 February 15: Sunquakes Produced by Flux Rope Eruption}.
\newblock {\em \apj}, {\bf 741}, L35.

\end{thebibliography}

\end{document}